\documentclass[aps,prd,twocolumn]{revtex4-1}

\usepackage {graphicx}
\usepackage {amsmath}
\usepackage {amsfonts}
\usepackage {amssymb}
\usepackage {mathrsfs}
\usepackage {bm}

\usepackage {epstopdf}

\newcommand {\be}{\begin{eqnarray}}
\newcommand {\ee}{\end{eqnarray}}

\begin{document}

\title{Isolated Vortex and Vortex Lattice in a Holographic $p$-wave Superconductor}



\author{James M. Murray}
\affiliation{Institute for Quantum Matter and Department of Physics and Astronomy, Johns Hopkins University, Baltimore, Maryland 21218, USA}
\author{Zlatko Te\v{s}anovi\'{c}}
\affiliation{Institute for Quantum Matter and Department of Physics and Astronomy, Johns Hopkins University, Baltimore, Maryland 21218, USA}


\date{\today}

\begin{abstract}
Using the holographic gauge-gravity duality, we find a solution for an isolated vortex and a vortex lattice in a 2+1 dimensional $p$-wave superconductor, which is described by the boundary theory dual to an $SU(2)$ gauge theory in 3+1 dimensional anti-de Sitter space. Both $p_x+ip_y$ and $p_x-ip_y$ components of the superconducting order parameter, as well as the effects of a magnetic field on these components, are considered. The isolated vortex solution is studied, and it is found that the two order parameter components have different amplitudes due to the time reversal symmetry breaking. The vortex lattice for large magnetic fields is also studied, where it is argued that only one order parameter component will be nonzero sufficiently close to the upper critical field. The upper critical field exhibits a characteristic upward curvature, reflecting the effects of field-induced correlations captured
by the holographic theory. The free energy is calculated perturbatively in this region of the phase diagram, and it is shown that the triangular vortex lattice is the thermodynamically preferred solution.
\end{abstract}


\maketitle


Recently the gauge-gravity duality \cite{maldacena98, witten98} has provided a new means by which to explore strongly interacting theories in condensed matter systems. One of the earliest and most successful of such models has been shown to exhibit the key properties of superconductivity: a phase transition at a critical temperature, where a spontaneous symmetry breaking of a $U(1)$ gauge symmetry in the bulk gravitational theory corresponds to a broken global $U(1)$ symmetry on the boundary, and the formation of a charged condensate. Gravity duals have so far been found for $s$-wave superconductors \cite{herzog08, hartnoll08} (in which the Cooper electron pairs have angular momentum $l=0$), as well as for $p$-wave ($l=1$) \cite{gubser08, roberts08, gubser08b} and $d$-wave ($l=2$) \cite{chen10, benini10} superconductors. The field of holographic superconductivity has since grown rapidly (see Ref.\ \cite{reviews} for reviews). Such a dual description provides a window through which we might hope to obtain insight into the properties and behaviors of superconductors and superfluids that defy description by more traditional approaches. 

The basic recipe for creating a holographic superconductor involves the introduction of a black hole and a charged scalar field into anti-de Sitter (AdS) spacetime in $d$+1 dimensions. According to the AdS/CFT correspondence, this theory is dual to a $d$-dimensional field theory that exists on the boundary of this space, where the boundary value of the bulk field is related to the expectation value of an operator, which is interpreted as the superconducting order parameter, in the boundary theory, and the temperature of the boundary theory is given by the Hawking temperature of the black hole. Below a critical temperature $T_c$ the field condenses, and the operator on the boundary acquires a nonzero expectation value, which corresponds to a superconducting phase transition.

The simplest case in the above scenario is that in which the electromagnetic and scalar fields are a function only of the radial AdS coordinate, having a boundary value which is spatially uniform. Such solutions were soon extended to include solutions for isolated vortices, which feature a spatially non-uniform order parameter, in $s$-wave superconductors \cite{montull09, albash09, domenech10} and vortex lattices in $s$-wave \cite{maeda10} and $d$-wave \cite{zeng10} superconductors. Extending these results to $p$-wave superconductors is of interest for a number of reasons. The order parameter in these systems has multiple components and breaks time-reversal symmetry, which leads to a richer set of possibilities than is possible in the simpler $s$-wave superconductors. In addition, the fact that the gravity dual is a $SU(2)$ gauge theory makes its development more straightforward than for $d$-wave superconductors, where the dual theory involves a spin-2 field in a gravitational background. The theory of holographic $p$-wave superconductors is also attractive because it has fewer free parameters than the charged scalar field theories that describe $s$- and $d$-wave superconductors.

The action proposed in Ref.\ \cite{gubser08b} to describe a $p$-wave holographic superconductor is
\be
\label{action}
S = \frac{1}{\kappa^2} \int d^4x \sqrt{-g} \left[ R + \frac{6}{L^2} - \frac{1}{4 q^2} (F^a_{\mu \nu})^2 \right],
\ee
where $\kappa$ is the gravitational coupling, $R$ is the Ricci scalar curvature, $L$ is the radius of AdS space and $F^a_{\mu \nu} = \partial_\mu A^a_\nu - \partial_\nu A^a_\mu + \epsilon^{abc} A^b_\mu A^c_\nu$ is the $SU(2)$ Yang-Mills field strength. The bulk gravitational theory is described by the AdS-Schwarzschild metric:
\be
ds^2 = \frac{L^2}{z^2} \left[ -g(z) dt^2 + \frac{dz^2}{g(z)} + dx^2 + dy^2 \right] ,
\ee
where $z$ is the radial AdS coordinate and $g(z) = 1-(z/z_0)^3$. The black hole horizon at $z=z_0$ is related to the Hawking temperature of the black hole, which is equal to the temperature of the boundary theory, by $z_0 = 3 / (4 \pi T)$. Our calculations will be performed in the probe limit, in which there is no back-reaction of the gauge field on the metric \cite{hartnoll08}.

As a starting point we consider the following ansatz for the gauge field:
\be
\label{ansatz}
\begin{split}
A =& \tau^3 ( \Phi dt + A^3_x dx + A^3_y dy )+ w_+ (\tau^1 dx + \tau^2 dy)
\\ & + w_- (\tau^1 dx - \tau^2 dy).
\end{split}
\ee
Here $\tau^a$ are the generators of $SU(2)$, which obey the relation $[\tau^a,\tau^b] = \epsilon^{abc} \tau^c$ and are related to the Pauli matrices by $\tau^a = \sigma^a / 2i$. Following Refs.\ \cite{gubser08, roberts08, gubser08b}, we interpret the $U(1)$ subgroup generated by $\tau^3$ as the group of electromagnetism, so that $\Phi (x,y,z)$ and $A^3_{x,y} (x,y,z)$ are the electromagnetic scalar and vector potentials, respectively. Because $\tau^3$ generates a rotation in the 1-2 ``plane," which is the analogue of a rotation in the complex plane in an ordinary Ginzburg-Landau theory of superconductivity, the (real) scalar fields $w_\pm (x,y,z)$ are charged under this $U(1)$, and they represent the amplitudes of the $p_x \pm i p_y$ components of the superconducting order parameter, respectively.

To study the case of an isolated vortex, we switch boundary coordinates from $(x,y)$ to $(r,\phi)$. In an ordinary Ginzburg-Landau theory with a complex order parameter, the vortex solution is found by replacing $\psi (r, \phi) \to e^{i n \phi} \psi (r)$, such that the phase changes by $2 \pi n$ as one goes around the vortex core, and $n$ is known as the winding number. By analogy, the vortex ansatz for the $p_x \pm i p_y$ superconductor is given by the replacements $w_\pm (r,\phi,z) \to \mathrm{exp}(2 n_\pm \phi \tau^3) w_\pm (r,z)$ in Eq.\ \eqref{ansatz}, where $n_\pm$ are the (integer) winding numbers for the two components of the superconducting order parameter. With this modification, the gauge field ansatz becomes
\be
\begin{split}
A^1_x &= w_+ (r,z) \cos(n_+ \phi) +  w_- (r,z) \cos (n_- \phi)
\\ A^1_y &= - w_+ (r,z) \sin(n_+ \phi) -  w_- (r,z) \sin (n_- \phi)
\\ A^2_x &= w_+ (r,z) \sin(n_+ \phi) -  w_- (r,z) \sin(n_- \phi)
\\ A^2_y &= w_+ (r,z) \cos(n_+ \phi) -  w_- (r,z) \cos (n_- \phi).
\end{split}
\ee
Furthermore, we assume that the electromagnetic scalar and vector potentials are rotationally symmetric and given by $\Phi(r,z)$ and $A^3_\phi (r,z)$, respectively.

The next step is to determine and numerically solve the equations of motion for this ansatz. The Yang-Mills equations are
\be
\label{ym}
0 = \frac{1}{\sqrt{-g}} \partial_\mu (\sqrt{-g}F^{a \mu \nu}) + \epsilon^{abc} A^b_\mu F^{c \mu \nu}.
\ee
To get the equation of motion for $w_+(r,z)$, we add the Yang-Mills equations \eqref{ym} for $(a,\nu) = (1,x)$ and $(a,\nu) = (2,y)$. We find that to obtain consistent equations, with $w_+$ independent of $\phi$, requires that the winding numbers are related by $n_- = n_+ + 2$. This relation between the winding numbers is also seen in the ordinary Ginzburg-Landau theory of $p$-wave superconductors \cite{heeb99}, and is ultimately due to the presence of mixed gradient terms such as $(D_x + iD_y)^2$ in the full equations of motion. Choosing $n_\pm = \mp 1$, which we expect corresponds to the lowest energy solution, yields
\be
\begin{split}
\label{eom1}
0 =& \partial_z (g \partial_z w_+) + \frac{1}{2r} \partial_r (r \partial_r w_+) - \frac{1}{2} \partial_r \left[ \frac{1}{r} \partial_r (r w_-) \right] 
\\ & - \frac{1}{2r} w_- \partial_r A^3_\phi - \frac{1}{r} A^3_\phi \partial_r w_- -\frac{1}{2r^2}(A^3_\phi)^2 w_- 
\\ & + \left[ \frac{\Phi^2}{g} + w_-^2 - w_+^2 - \frac{(A^3_\phi - 1)^2}{2r^2} \right] w_+.
\end{split}
\ee
Similarly, the equations of motion for $w_-$ and the gauge fields are
\begin{align}
\begin{split}
\label{eom2}
0 =& \partial_z (g \partial_z w_-) + \frac{1}{2r} \partial_r (r \partial_r w_-) - \frac{1}{2} \partial_r \left[ \frac{1}{r} \partial_r (r w_+) \right] 
\\ & + \frac{1}{2r} w_+ \partial_r A^3_\phi + \frac{1}{r} A^3_\phi \partial_r w_+ -\frac{1}{2r^2}(A^3_\phi)^2 w_+ 
\\ & + \left[ \frac{\Phi^2}{g} + w_+^2 - w_-^2 - \frac{(A^3_\phi + 1)^2}{2r^2} \right] w_-
\end{split}
\\ 0 =& \partial^2_z \Phi + \frac{1}{r g} \partial_r (r \partial_r \Phi) - \frac{2}{g}(w_+^2 + w_-^2)\Phi \\
\begin{split}
0 =& \partial_z (g \partial_z A^3_\phi) + r \partial_r \left( \frac{1}{r} \partial_r A^3_\phi \right) + r \partial_r (w_-^2 - w_+^2) 
\\ &+ (w_+ + w_-) \partial_r A^3_\phi + r (w_+ + w_-)(w_-^2 - w_+^2).
\end{split}
\end{align}

These equations can be solved numerically, subject to appropriate boundary conditions. At the boundary of AdS at $z=0$, the fields have the limiting forms
\be
\begin{split}
w_\pm =& \left< {\cal O}_\pm \right> z + \ldots
\\ \Phi =& \mu - \rho z + \ldots
\\ B (r) =& \frac{1}{r} \partial_r A^3_\phi, 
\end{split}
\ee
where $\sqrt{ \left<{\cal O}_\pm \right> }$ are interpreted as the two components of the superconducting order parameter (the square root is necessary since ${\cal O}_\pm$ has mass dimension 2, whereas the superconducting order parameter should have mass dimension 1), $\mu$ is the chemical potential, $\rho$ is the charge density, and $B(r)$ is the magnetic field. In our numerical solution, we specify the value of $\mu$, as well as the conditions $w_\pm (z=0)=0$ and $\partial_z A^3_\phi = 0$. As discussed in Ref.\ \cite{domenech10}, this last condition is a Neumann boundary condition, which, unlike the more commonly used Dirichlet condition, allows for the presence of a dynamical gauge field in the boundary theory, and also, according to the AdS/CFT dictionary, leads to vanishing of the current operator in the boundary theory. At the horizon ($z=z_0$), we require $\Phi = 0$ and $A^3_\phi$ be regular. At the vortex core ($r=0$), the boundary conditions are $w_\pm = 0$, $A^3_\phi = 0$ and $\partial_r \Phi = 0$. Finally, far from the vortex core at the edge of our solution domain ($r=R$), we require $\partial_r w_\pm = 0$, $\partial_r \Phi = 0$ and $A^3_\phi = 1$. This last condition ensures that there is one quantum of magnetic flux passing through the vortex region \cite{domenech10}, with $\int d^2 r B(r) = 2 \pi$. To obtain our numerical solutions we have used the COMSOL 3.4 package \cite{comsol}.
\begin{figure}
\includegraphics[width=0.4\textwidth]{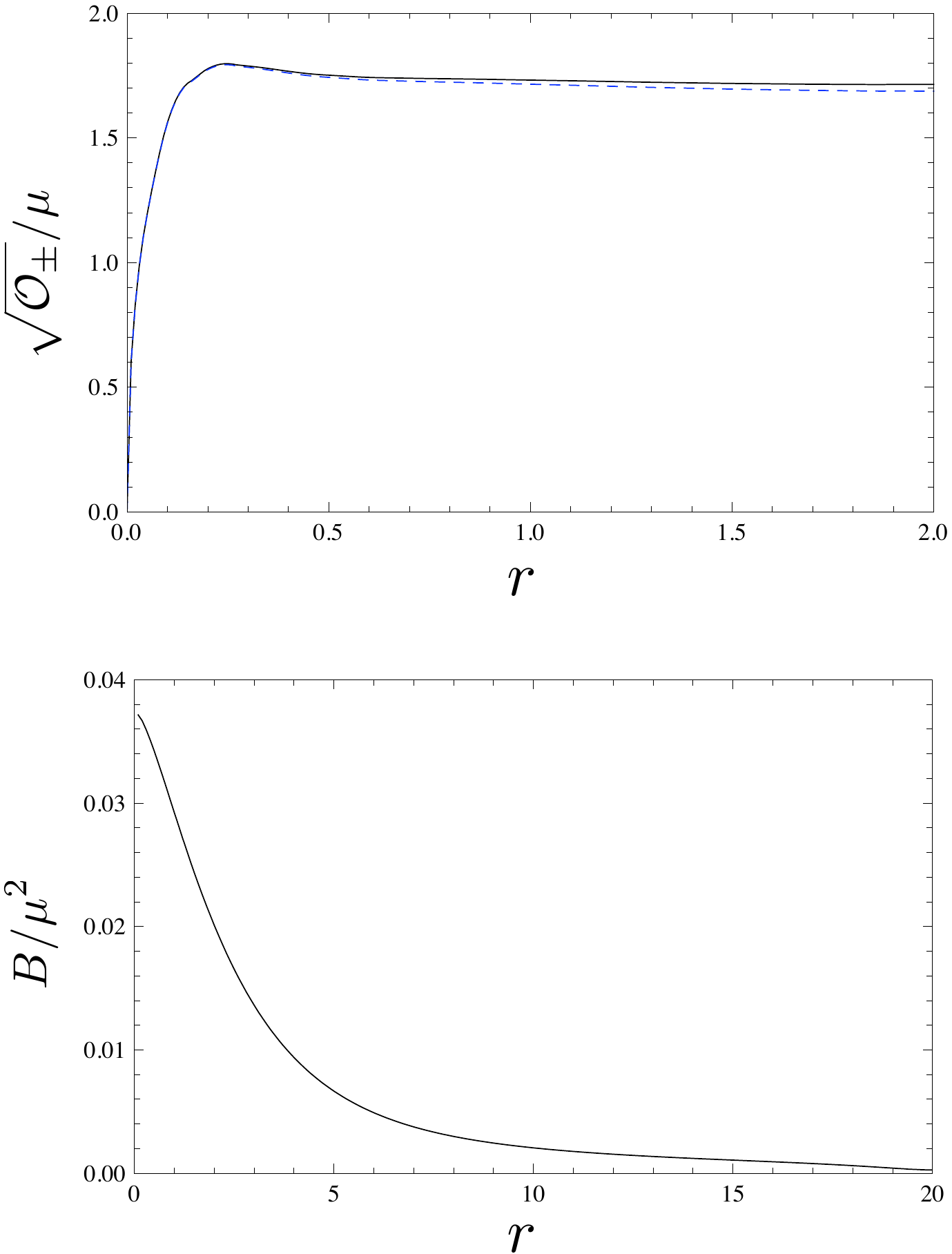}
\caption{(Top) Spatial profile of the $w_+$ (solid) and $w_-$ (dashed) components of the superconducting order parameter for an isolated vortex in a magnetic field at temperature $T/\mu=0.032$. (Bottom) Magnetic field profile for the same vortex configuration.
\label{vortex_profile}}
\end{figure}

Fig.\ \ref{vortex_profile} shows the spatial profile of the two components of the superconducting order parameter for an isolated vortex, through which a single quantum of magnetic flux penetrates the superconductor. It can be seen that, as the order parameter approaches its bulk value far from the vortex core, the $w_-$ component has a slightly smaller amplitude than the $w_+$ component. This is a consequence of the breaking of time-reversal symmetry, and the fact that the two components do not couple to the external field in the same way. The difference between the two components will grow as the field increases, however vortices will tend to proliferate at higher fields, so that the picture of an isolated vortex eventually ceases to be valid. The hump at $r \sim 0.2$ is an interesting feature that appears to be present at all temperatures. It may arise from the fact that the gradient terms in our theory are different from those in the usual Ginzburg-Landau theory, where such a hump is generally not present \cite{heeb99}.

The lower part of Fig.\ \ref{vortex_profile} shows the profile of the magnetic field near the vortex core. The exponential decay of the field with distance from the vortex core is a general property of superconductors and is also seen in the conventional Ginzburg-Landau theory. It is interesting to compare the size of the vortex core (the so-called ``coherence length") $\xi \sim 0.1$ to the penetration depth of the magnetic field $\lambda \sim 3$ \cite{lambda_note}. The ratio of these quantities defines the Ginzburg-Landau parameter $\kappa \equiv \lambda / \xi \sim 30$. The fact that $\kappa \gg 1$ means that the holographic $p$-wave superconductors are strongly type II, and are therefore expected to exhibit a vortex lattice solution near an upper critical magnetic field $H_{c2}$. Most of the superconductors that attract widespread theoretical interest, including the high-temperature cuprate superconductors, fall in this regime.  

We begin our discussion of the superconducting properties near the upper critical field by considering the case of an ordinary, non-holographic $p$-wave superconductor in a magnetic field. In this case the ground state that is realized is known to depend on the coefficients of the kinetic terms in the free energy. For a two-dimensional superconductor with two complex components, the kinetic terms allowed by symmetry are $K_1 (D_i \eta_j)^* D_i \eta_j$, $K_2 (D_i \eta_i)^* D_j \eta_j$ and $K_3 (D_i \eta_j)^* D_j \eta_i$, and the ground state that is realized depends on the values of these coefficients \cite{sundaram89}. For $(K_2+K_3)/K_1 < 0$ the free energy is not bounded from below and the theory is unstable. (This is true at least for the linearized version of the theory. The theory may still be stable when higher order terms are included.) In the stable region, one of two possible ground states is realized. For $(K_2+K_3)/K_1 > 0$ and $K_2-K_3 > (K_2+K_3)^2/(2K_1+K_2+K_3)$ only one order parameter component is nonzero (e.g.\ $\eta_+ \sim \eta_1 + i \eta_2$ is nonzero if the field is in the $+ \hat z$ direction), and this component is in the lowest ($n=0$) Landau level. Such a state shall be denoted as $|0\rangle_+$. On the other hand, for $K_2-K_3 < (K_2+K_3)^2/(2K_1+K_2+K_3)$, both order parameter components are nonzero, and there is a mixture of $n=0$ and $n=2$ Landau levels (e.g. the state is of the form \ $c_+ |2\rangle_+ + c_- |0\rangle_-$ if the field is in the $+ \hat z$ direction, where the coefficients will generally depend on temperature and field). Following Ref.\ \cite{sundaram89}, we call these, respectively, the $A$ and $U$ phases. Writing out the Yang-Mills Lagrangian from Eq.\ \eqref{action} in terms of the fields $w_\pm$, it can be shown that the coefficients in our theory are $K_1 = 1$, $K_2 = 0$ and $K_3 = -1$, so it appears that we are on the boundary between the stable and unstable regions, and also--if we assume that the phase boundary remains unchanged for the holographic superconductor--on the boundary between the $A$ and $U$ phases described above. One can imagine changing the coefficients of these gradient terms by hand, thereby moving away from this critical point, but in that case we would no longer be dealing with pure Yang-Mills theory, which has the attractive feature of having no adjustable parameters. 

Of course, the extent to which these results are relevant for holographic superconductors ought to be questioned. Since our ansatz leads to a Lagrangian in which there are no terms coupling gradients in the $z$ direction to those in the $xy$ plane, the criterion for stability should remain the same as in the non-holographic case. Also, by assuming that one order parameter component vanishes and employing separation of variables, it is shown below that the $A$ phase is a solution in the stable region of the phase diagram for the holographic case. Separation of variables, however, cannot be used to obtain the more complicated $U$ phase described above. It is possible that there is an analogue of this phase in the holographic superconductor, but because the $U$ phase in the non-holographic case is a complicated function involving multiple Landau levels, we expect that the same would be true in the holographic case. Characterizing such a state would most likely involve minimizing numerically the free energy of variational wavefunctions, and the problem of determining the exact form of such a state and comparing its energy with that of the $A$ phase is an interesting problem left for future study. It is important to note, however, that the criterion given above for distinguishing the $A$ and $U$ phases will not necessarily hold in the holographic case, so it is possible that the $A$ phase is in fact the unambiguous ground state for the particular Lagrangian that defines our theory. We therefore proceed pragmatically, assuming stability and the existence of a state with $w_-=0$ and following the original approach of Abrikosov \cite{abrikosov57}, which was also used in Refs.\ \cite{maeda10, zeng10}. (If $B<0$, the following discussion holds if we replace $w_+ \to w_-$.) 

Since there will not be rotational symmetry for the vortex lattice as there was for the isolated vortex, we once again take all fields to be functions of all three spatial variables, with $A^a_\mu = A^a_\mu (x,y,z)$. In order to make notation more transparent and to allow for easier comparison with the existing literature on superconductivity, we now switch to more conventional notation in which the superconducting order parameters are represented by two complex scalar fields. Letting
\be
\begin{split}
\eta_{1,2} = A^1_{x,y} + iA^2_{x,y}
\\ \eta_\pm = \frac{1}{\sqrt 2} (\eta_1 \pm i \eta_2),
\end{split}
\ee
the Yang-Mills part of the action in Eq.\ \eqref{action} can be expressed as
\be
\begin{split}
\label{gl_action}
S^{YM} & = \frac{1}{2 q^2 \kappa^2} \int d^4x \bigg[ (\partial_z \Phi)^2 + \frac{1}{g} (\nabla \Phi )^2 + \frac{\Phi^2}{g} |\eta_+|^2 
\\ & - g \left[ |\partial_z \eta_+|^2 + (\partial_z {\bf A})^2 \right] - \frac{1}{2} |D_x \eta_+ |^2  
\\ &  - \frac{1}{2} |D_y \eta_+ |^2 - \left( \partial_x A_y - \partial_y A_x + \frac{1}{2} |\eta_+|^2 \right)^2
\\ & + \frac{1}{2i} \left[ D_x \eta_+ ( D_y \eta_+)^* - ( D_x \eta_+)^* D_y \eta_+ \right]  \bigg] ,
\end{split}
\ee
where we have assumed that we are in the $A$ phase described above and sufficiently close to the upper critical field that we can set the second order parameter component $\eta_- = 0$. Here and in what follows, the gradient operator is defined as $\nabla \equiv (\partial_x , \partial_y )$, and to simplify notation we have let $A^3_i \to A_i$. 

As was done in Refs.\ \cite{maeda10, zeng10}, near the upper critical field, we can define $\epsilon \equiv (H_{c2} - H)/H_{c2}$ and expand the fields:
\be
\begin{split}
\Phi(x,y,z) &= \Phi^{(0)} + \epsilon \Phi^{(1)} + {\cal O} (\epsilon^2)
\\ A_{x,y} (x,y,z) &= A_{x,y}^{(0)} + \epsilon A_{x,y}^{(1)} + {\cal O} (\epsilon^2)
\\ \eta_+ (x,y,z) &= \epsilon^{1/2} \eta_+^{(1)} + \epsilon^{3/2} \eta_+^{(2)} + {\cal O} (\epsilon^{5/2}).
\end{split}
\ee
To leading order near $H_{c2}$, the electromagnetic fields are $\Phi = \mu (1-z/z_0)$, $A^3_y = x H_{c2}$ and $A^3_x = 0$. The higher order terms take into account the backreaction of the bosonic field on the electromagnetic fields. The magnetic field in the boundary theory is $H \equiv (\partial_x A_y - \partial_y A_x) |_{z=0}$. Taking the equation of motion for $\eta_+$ from Eq.\eqref{gl_action} and letting $\eta_+ (x,y,z) = m_+ (x,z;p) e^{ipy}$ gives
\be
\begin{split}
\label{eom_m}
0 =& \partial_z (g  \partial_z m_+) + \frac{1}{2} \partial_x^2 m_+ 
\\ & + \left[ \frac{\Phi^2}{g} - \frac{1}{2} (H_{c2} x + p )^2 - \frac{3}{2} H_{c2} \right] m_+ ,
\end{split}
\ee
where we have neglected the term $\sim |m_+|^2 m_+$, since $m_+$ is small near $H_{c2}$. We note here that the vortex lattice solution for the $p$-wave superconductor was not obtained in Ref.\ \cite{zeng10} because the {\it gauge field} ansatz in that paper contained a complex phase factor, so that $A_\mu \sim e^{i p y}$. In our work, it is the complex bosonic field $\eta_+$ which is given the phase factor, which, in the language of the original gauge fields, corresponds to $A_\mu \sim e^{2 p y \tau^3}$. This is necessary to obtain the term $\sim (H_{c2}x+p)^2$ in Eq.\ \eqref{eom_m}. The distinction is important because, as we described above, the gauge field in this theory is real, with the role of the real and imaginary parts of the usual Ginzburg-Landau order parameter being played here by the $\tau^1$ and $\tau^2$ directions in $SU(2)$ space.

\begin{figure}
\includegraphics[width=0.4\textwidth]{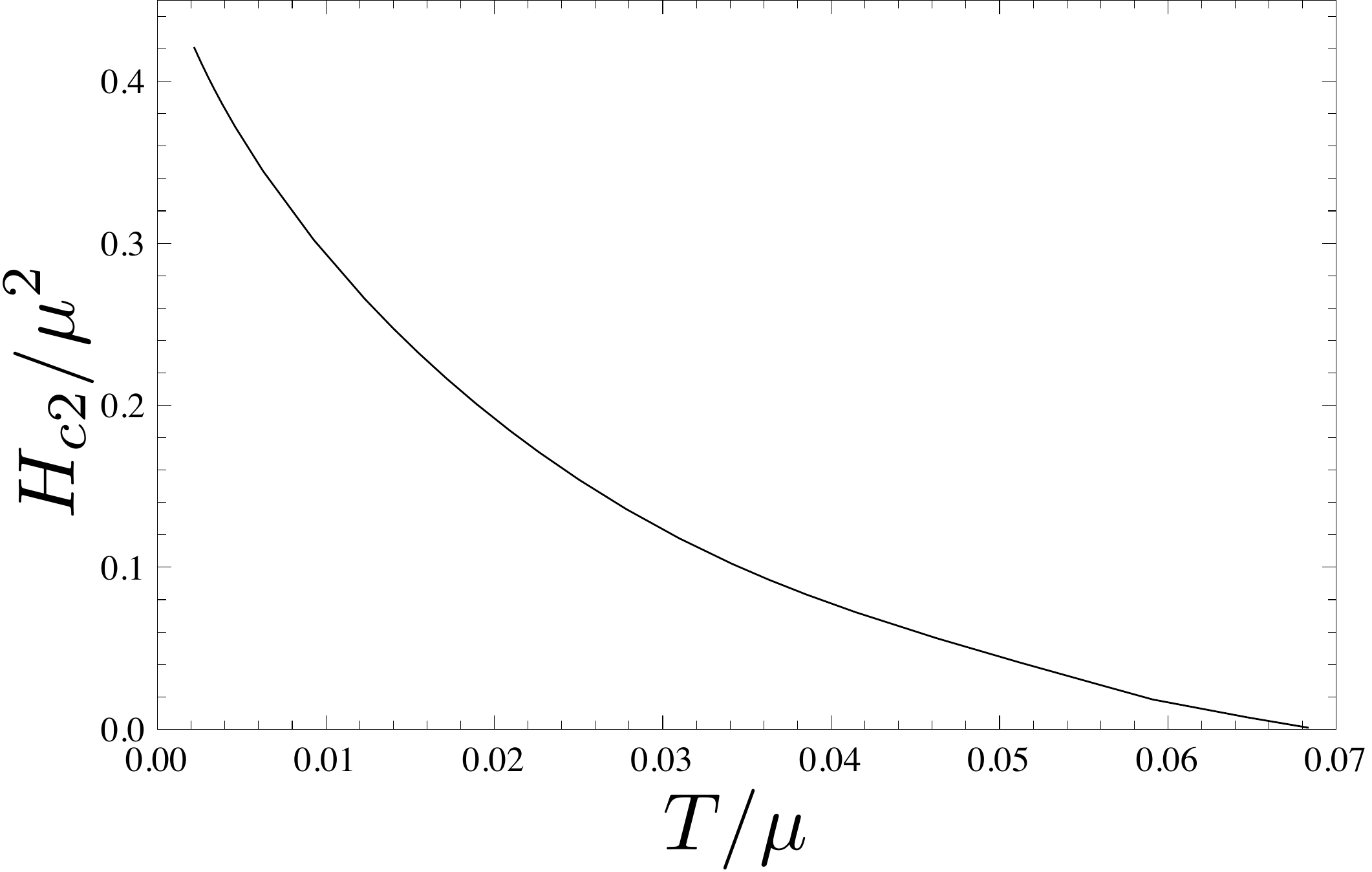}
\caption{Upper critical magnetic field for the $w_+$ component of the superconducting order parameter. The lower left part of the phase diagram is the superconducting region, and the upper right is the normal state. Note the upward curvature, which is
a direct consequence of the $z$-dependence in Eq.\ \eqref{rho}. 
\label{bc2}}
\end{figure}

Taking advantage of the linearity of Eq.\ \eqref{eom_m}, we can again use separation of variables, letting $m_+ (x,z;p) = \rho(z)\gamma(x;p)$. We then obtain the following eigenvalue equations:
\begin{align}
0 =& -\partial_X^2 \gamma_n + X^2 \gamma_n - \lambda_n \gamma_n
\\ 0 =& \partial_z (g \partial_z \rho_n) + \left[ \frac{\mu^2}{g} \left( 1-\frac{z}{z_0} \right)^2 - \frac{H_{c2}}{2} (\lambda_n + 3) \right] \rho_n,
\label{rho}
\end{align}
where $X \equiv \sqrt{H_{c2}} (x+p/H_{c2})$. The first of these is just a harmonic oscillator equation, which is solved by the Hermite polynomials:
\be
\gamma_n (x;p) = e^{-X^2/2}H_n(X).
\ee
Here $n = 0,1,2,\ldots$ denotes the Landau energy level, and the corresponding eigenvalues are given by $\lambda_n = 2n+1$. The Abrikosov vortex lattice is given by a superposition of the lowest energy ($n=0$) solutions:
\be
m_+ (x,y,z) = \rho_0 (z) \sum_j c_j e^{i p_j y} \gamma_0 (x;p_j),
\ee
where the $c_j$ are coefficients that determine the structure of the vortex lattice. As shown in Refs.\ \cite{maeda10,zeng10,ge10}, the upper critical field $H_{c2}$ can be calculated at a given temperature by finding the highest field at which Eq.\ \eqref{rho} has a non-vanishing solution, indicating the presence of a superconducting condensate. The resulting phase diagram is shown in Fig.\ \ref{bc2}. We alert the reader to the characteristic upward curvature of $H_{c2}$. This curvature, stemming from Eq.\ (\ref{rho}),
is intrinsic to our theory and is thus reflective of the effects of field-induced correlations captured within the holographic approach.

We next investigate the nature of the vortex lattice solution near $H_{c2}$. It is well known that the free energy in an ordinary ($s$-wave) Ginzburg-Landau theory is minimized when the vortex cores form a triangular lattice, and this has also been shown to be the case for a holographic $s$-wave superconductor \cite{maeda10}. Here we follow a similar approach to find the configuration that minimizes the free energy of the holographic $p$-wave superconductor. While our analysis is complicated somewhat compared to the $s$-wave case by the quartic term and the many gradient terms that appear in Eq.\ \eqref{gl_action}, we shall find that--just as in the $s$-wave case--the triangular vortex lattice minimizes the free energy.

Since all quantities are time-independent, the free energy is given by $\Omega = - S^{YM}_{\mathrm{OS}}/\int dt$, where $S^{YM}_{\mathrm{OS}}$ is the action evaluated with the fields $\eta_+$, $A_{x,y}$ and $\Phi$ on shell. The equation of motion for $\eta_+$ following from Eq.\ \eqref{gl_action} is
\be
\begin{split}
\label{eta_eom}
0 = & \partial_z ( g \partial_z \eta_+) + \frac{1}{2} (D_x + iD_y)(D_x - iD_y) \eta_+
\\ & + \frac{\Phi^2}{g} \eta_+ + (\partial_x A_y - \partial_y A_x) \eta_+ - \frac{1}{2} |\eta_+|^2 \eta_+ .
\end{split}
\ee
Multiplying this equation by $\eta_+^*$ and integrating over space, combining this result with Eq.\ \eqref{gl_action} gives the action with $\eta_+$ evaluated on shell:
\be
\label{omega}
\begin{split}
S^{YM} (\bar \eta_+) = & \frac{-1}{ 2 q^2 \kappa^2} \int d^4x \bigg[ - (\partial_z \Phi)^2 - \frac{1}{g} (\nabla \Phi)^2 
\\ & + g (\partial_z {\bf A})^2 + (\partial_x A_y - \partial_y A_x )^2 - \frac{1}{4} |\bar \eta_+|^4 \bigg],
\end{split}
\ee
where integration by parts has been used, and a bar denotes fields evaluated on shell. We assume that the scalar field $\eta_+$ has compact support in the $(x,y)$ coordinates, so that the contributions from the boundaries $x = const.$ and $y=const.$ vanish when we integrate by parts. Furthermore, there is no boundary contribution from the horizon due to the regularity condition, and none from the AdS boundary at $z=0$ due to the boundary condition $\eta_+(z=0)=0$.

Similarly, the on-shell conditions for the fields $\Phi$ and $A_i$ can be calculated from Eq.\ \eqref{omega} and the result substituted back into the action, yielding the action with all fields evaluated on shell
\be
\begin{split}
S^{YM}_{\mathrm{OS}} = \frac{1}{ 2 q^2 \kappa^2} \bigg[ & \frac{1}{4} \int d^4x g |\bar \eta_+|^4 
\\ & + \int d^3x \bar A_i \partial_z \bar A_i \bigg|_{z=0} \bigg].
\end{split}
\ee
Here there is no boundary contribution from the $\Phi$ terms upon integration by parts due to the compact support in the  $(x,y)$ coordinates and the boundary condition $\Phi (z=0) = 0$. As discussed in Ref.\ \cite{maeda10}, the boundary contributions from the gauge fields at the horizon can also be ignored since they are independent of the field $\eta_+$, and our interest here is in finding the configuration of $\eta_+$ that minimizes the free energy. The boundary term at $z=0$ for the field $A_i$ does not in general vanish, however; and, according the AdS/CFT dictionary, the expectation value of the current operator in the boundary theory is proportional to $\partial_z A_i$. 

The leading nonzero correction to the action is at ${\cal O}(\epsilon^2)$, so the free energy can be expressed as $\Omega \approx \Omega^{(0)} + \epsilon^2 \Omega^{(2)}$, with
\be
\label{omega2}
\begin{split}
\Omega^{(2)} =  \frac{-1}{ 2 q^2 \kappa^2} \int d{\bf x} \bigg[ & \frac{1}{4} \int dz g | \eta^{(1)}_+|^4 
\\ & + A^{(1)}_i \partial_z A^{(1)}_i \bigg|_{z=0} \bigg].
\end{split}
\ee
For simplicity we no longer use the bar to denote fields that are on shell. In order to evaluate Eq.\ \eqref{omega2}, we need to obtain an expression for $A_i^{(1)}$. At ${\cal O} (\epsilon)$, the equation of motion following from Eq.\ \eqref{gl_action} is
\be
\label{a_eom}
[ \partial_z (g \partial_z) + 2 \nabla^2 ] A^{(1)}_i = -j_i^{(1)} + \frac{1}{2} \epsilon_{ij} \partial_j |\eta_+^{(1)}|^2,
\ee
where we have defined
\be
\label{current}
j_i^{(1)} \equiv \frac{1}{2i} \left[ \eta_+^{(1)*} D_i^{(0)} \eta_+^{(1)} - (D_i^{(0)} \eta_+^{(1)})^* \eta_+^{(1)} \right] .
\ee
We have also chosen the gauge condition $\partial_i A_i = 0$ and applied it in deriving Eq.\ \eqref{a_eom}. It is important to note that Eq.\ \eqref{current} describes currents in the bulk theory, and is distinct from the boundary current operator alluded to above. Furthermore, since $\eta_+ (z,{\bf x}) \sim \gamma({\bf x})$, which describes the ground state of a simple harmonic oscillator, it is a simple matter to show \cite{degennes89} that $j_i^{(1)} = - \frac{1}{2} \epsilon_{ij} \partial_j |\eta_+^{(1)}|^2$. The antisymmetric symbol satisfies $\epsilon_{ji} = -\epsilon_{ij}$ and $\epsilon_{12}=1$.

The solution to Eq.\ \eqref{a_eom} can be expressed as
\be
\begin{split}
\label{a1}
A_i^{(1)}(z,{\bf x}) & = a_i ({\bf x})
\\ & - \int dz' d{\bf x}' G_B(z,z'; {\bf x} - {\bf  x'}) j_i^{(1)}(z',{\bf x'}),
\end{split}
\ee
where $a_i ({\bf x})$ is the homogeneous part of the solution satisfying $\epsilon_{ij}\partial_i a_j = - H_{c2}$, and the Greens function $G_B$ satisfies
\be
\begin{split}
\label{gb}
[ \frac{1}{2} \partial_z (g \partial_z) + \nabla^2 ] G_B & (z,z'; {\bf x} - {\bf  x'}) 
\\ & = - \delta(z-z') \delta({\bf x}-{\bf x'})
\\ G_B (z=0, {\bf x}) = \lim_{z \to z_0} & g(z) G_B (z,z'; {\bf x}) = 0.
\end{split}
\ee 

We can now use Eq.\ \eqref{a1} to evaluate the second term in Eq.\ \eqref{omega2}:
\be
\begin{split}
\label{bdy_int}
& \int d{\bf x} A_i^{(1)} \partial_z A_i
\\ & = \int dz' d{\bf x'} d{\bf x} a_i ({\bf x})  \partial_z G_B (z,z'; {\bf x}-{\bf x'}) j_i^{(1)}(z',{\bf x'})
\\ & = \frac{ H_{c2}}{2}  \int dz' d{\bf x'} d{\bf x} \partial_z G_B (z,z'; {\bf x}-{\bf x'}) |\eta_+^{(1)}(z',{\bf x'})|^2,
\end{split}
\ee
where in the second equality integration by parts and $\nabla G_B (z,z'; {\bf x}-{\bf x'}) = -\nabla' G_B (z,z'; {\bf x}-{\bf x'})$ were used. From the boundary condition in Eq.\ \eqref{gb}, the integral of $G_B$ can be performed:
\be
\int d{\bf x}G_B (z,z'; {\bf x}) = \int_0^{\mathrm{min}(z,z')} \frac{dz''}{g(z'')}.
\ee
Using this result along with Eq.\ \eqref{bdy_int}, we obtain the following expression for the free energy:
\be
\label{omega_2'}
\Omega^{(2)} =  \frac{-1}{ 4 q^2 \kappa^2} \int dz d{\bf x} \bigg[ & \frac{1}{2} g | \eta^{(1)}_+|^4 + H_{c2}  | \eta^{(1)}_+|^2 \bigg].
\ee

As pointed out in Ref.\ \cite{maeda10}, however, Eq.\ \eqref{omega_2'} is not our final result, since it depends on the normalization of $\eta_+^{(1)}$. This ambiguity in normalization is resolved by considering nonlinearity. From Eq.\ \eqref{eta_eom}, the equation of motion for $\eta_+^{(1)}$ is $\hat {\cal L} \eta_+^{(1)} = 0$, where we have defined the differential operator
\be
\begin{split}
\hat {\cal L} \equiv & \partial_z (g \partial_z) + \frac{1}{2}(D_x + iD_y)^{(0)}(D_x - iD_y)^{(0)} 
\\ & + \frac{(\Phi^{(0)})^2}{g} + H_{c2}.
\end{split}
\ee
Also following from Eq.\ \eqref{eta_eom} is the equation of motion for $\eta_+^{(2)}$, which can be expressed as $\hat {\cal L} \eta_+^{(2)} = J$, where
\be
\begin{split}
J \equiv & \frac{i}{2}(A_x + iA_y)^{(1)}(D_x -iD_y)^{(0)} \eta_+^{(1)} 
\\ & + \frac{i}{2}(D_x + iD_y)^{(0)} \left[ (A_x - iA_y)^{(1)} \eta_+^{(1)} \right] 
\\ & +\left[ (\partial_x A_y - \partial_y A_x)^{(1)} - \frac{2}{g} \Phi^{(0)} \Phi^{(1)} + \frac{1}{2} |\eta_+^{(1)}|^2 \right] \eta_+^{(1)}.
\end{split}
\ee
With this equation of motion, we obtain the following identity:
\be
\begin{split}
\label{oc}
0 &= \int d^3 x \left[  \eta_+^{(1)*} J - \eta_+^{(1)*} \hat {\cal L} \eta_+^{(2)} \right]
\\ &=  \int d^3 x \left[ \eta_+^{(1)*} J - \left( \hat {\cal L} \eta_+^{(1)} \right)^* \eta_+^{(2)} \right]
\\ &=  \int d^3 x  \eta_+^{(1)*} J,
\end{split}
\ee
where integration by parts along with the boundary condition $\eta_+ (z=0) = 0$ was used to obtain the second line. Again using integration by parts and $j_i^{(1)} = - \frac{1}{2} \epsilon_{ij} \partial_j |\eta_+^{(1)}|^2$ in Eq.\ \eqref{oc}, we obtain the condition
\be
\begin{split}
\label{oc2}
0 = & \int d^3x  \bigg[ \frac{1}{4} |\eta_+^{(1)}|^4 
\\ & - \left( \partial_x A^{(1)}_y - \partial_y A^{(1)}_x + \frac{2}{g} \Phi^{(0)} \Phi^{(1)} \right)  |\eta_+^{(1)}|^2 \bigg].
\end{split}
\ee

In order to evaluate Eq.\ \eqref{oc2} we must determine the form of $\Phi^{(1)}$. At ${\cal O}(\epsilon)$, the equation of motion for $\Phi$ is
\be
( g \partial_z^2 + \nabla^2 ) \Phi^{(1)} =  |\eta_+^{(1)}|^2 \Phi^{(0)}.
\ee
This equation has solution
\be
\label{phi1}
\begin{split}
\Phi^{(1)} (z,{\bf x}) = \int d{\bf x'} \int dz' & \frac{ \Phi^{(0)}(z') }{g(z')} G_t (z,z';{\bf x} - {\bf x'}) 
\\ & \times |\eta_+^{(1)} (z',{\bf x'})|^2,
\end{split}
\ee
where $G_t (z,z';{\bf x} - {\bf x'})$ is the Green function satisfying
\be
\begin{split}
\label{gt}
(g \partial_z^2 + \nabla^2) G_t (z,z';{\bf x} - {\bf x'}) =
\\ -g(z) \delta (z-z') \delta^{(2)} ({\bf x}-{\bf x'}).
\end{split}
\ee
In order to make further progress, we expand the Greens functions from Eqs.\ \eqref{gb} and \eqref{gt} in a basis of eigenfunctions:
\be
\begin{split}
\label{gt2}
G_t (z,z';{\bf x}) = \sum_\lambda \xi_\lambda (z) \xi^\dagger_\lambda (z') G_2 ({\bf x}, \lambda)
\\ -g(z) \partial_z^2 \xi_\lambda (z) = \lambda \xi_\lambda (z)
\\ \xi_\lambda (0) = 0 = \xi_\lambda (z_0),
\end{split}
\ee
and
\be
\begin{split}
\label{gb2}
G_B (z,z';{\bf x}) = \sum_\lambda \chi_\lambda (z) \chi^\dagger_\lambda (z') G_2 ({\bf x}, \lambda)
\\ - \frac{1}{2} \partial_z \left[ g(z) \partial_z \chi_\lambda (z) \right] = \lambda \chi_\lambda (z)
\\ \chi_\lambda (z=0) = \lim_{z\to z_0}g(z) \chi ' (z) = 0.
\end{split}
\ee
where the Greens function $G_2$ satisfies
\be
\label{g2}
(\nabla^2 - \lambda) G_2 ({\bf x},\lambda) = -\delta ({\bf x}).
\ee

Due to Eqs.\ \eqref{a1} and \eqref{phi1}, the free energy in this theory takes a nonlocal form, as opposed to the usual, non-holographic Ginzburg-Landau theory, which is completely local. This is due to the fact that the Ginzburg-Landau theory is a low energy effective expansion, whereas the AdS theory presented here retains the physics from all energy scales \cite{maeda10}. To get a local effective theory, we recognize that, in the long wavelength limit, $G_2 ({\bf x},\lambda)$ decays much more quickly than $|\eta_+^{(1)} ({\bf x})|^2$, so we can approximate
\be
\label{local}
\int d {\bf x'} G_2 ({\bf x} - {\bf x'}, \lambda) |\gamma ({\bf x'})|^2 \approx \frac{|\gamma ({\bf x})|^2}{\lambda}.
\ee 

Using Eqs.\ \eqref{a1} and \eqref{phi1}, we can now give an explicit form of the condition Eq.\ \eqref{oc2} in the long wavelength limit:
\be
\begin{split}
\label{oc3}
0 & = \int d^3 x \bigg[ \left( \frac{1}{4}  - \frac{2}{g}\Phi^{(0)} \Phi^{(1)} \right)  |\eta_+^{(1)}|^4 - H_{c2}  |\eta_+^{(1)}|^2
\\ &-  |\eta_+^{(1)}|^2 \nabla^2 \int d^3x' G_B(z,z'; {\bf x} - {\bf x'})  |\eta_+^{(1)}(z',{\bf x'})|^2  \bigg]
\\ & \approx \int d^3 x \bigg[ \left( \frac{\rho^4}{4}  - \frac{\alpha(z) \rho^2}{2} \right)  |\gamma|^4 - H_{c2} \rho^2 |\gamma|^2 \bigg]
\end{split}
\ee
To obtain the first equality we have again used integration by parts and $\nabla' G_B(z,z'; {\bf x} - {\bf x'})  = -\nabla G_B(z,z'; {\bf x} - {\bf x'}) $, and the second equality gives the approximate form in the long wavelength limit, using Eqs.\ \eqref{gt2}-\eqref{g2}. We have also defined
\be
\alpha (z) \equiv \frac{4 \Phi^{(0)}}{g} \sum_\lambda \frac{ \xi_\lambda}{\lambda} \int dz' \frac{\Phi^{(0)}(z')}{g(z')} \rho^2 (z') \xi^\dagger_\lambda (z').
\ee

By combining Eqs.\ \eqref{oc3} and \eqref{omega_2'}, we can now give an approximate, local expression for the free energy density that is independent of the normalization of the order parameter:
\be
\begin{split}
\label{free_energy}
\frac{\Omega}{V} &= \frac{1}{V} \left[ \Omega^{(0)} + \epsilon^2 \Omega^{(2)} + \ldots \right]
\\ & \approx  \frac{\Omega^{(0)}}{V} - \frac{2 (H_{c2}-H)^2 \left< \rho^2 \right> ^2 \left< 2 \rho^4 - \alpha \rho^2 \right>}{q^2 \kappa^2 \beta \left< \rho^4 - \alpha \rho^2 \right> ^2}, 
\end{split}
\ee
where $\left< \ldots \right>$ denotes spatial average, and the Abrikosov parameter is given by
\be
\beta \equiv \frac{\left< |\gamma |^4 \right>}{\left< |\gamma |^2 \right> ^2}.
\ee 
Since the free energy density in Eq.\ \eqref{free_energy} is negative, minimizing the free energy corresponds to minimizing $\beta$. It is well known that the vortex lattice distribution that minimizes $\beta$ is the triangular vortex lattice, for which $\beta = 1.159$. This was also the result found for the holographic $s$-wave superconductor \cite{maeda10}.

In conclusion, we have shown the existence of a vortex solution in a holographic $p$-wave superconductor at low magnetic fields, as well as a vortex lattice solution near the upper critical magnetic field, $H_{c2}$. $H_{c2}$ exhibits a characteristic upward curvature, intrinsic to our theory, which reflects the effects of field-induced correlations captured by the holographic approach. The free energy was found to be minimized by the triangular vortex lattice. In the future it would be interesting to extend this theory to a BCS-like theory of fermions in AdS \cite{hartman10}, which would give insight into the possible types of $p$-wave pairing in holographic superconductors, as well as the tantalizing possibility of Majorana fermions, which are known to exist as bound states in the vortex cores of chiral $p$-wave superconductors.

\begin{acknowledgments}
We thank D.\ E.\ Kaplan, A.\ Salvio, I.\ Tolfree and Y.\ Wan for useful discussions. This work was supported by the Johns Hopkins-Princeton Institute for Quantum Matter, under Award No.\ DE-FG02-08ER46544 by the U.S. Department of Energy, Office of Basic Energy Sciences, Division of Materials Sciences and Engineering.
\end{acknowledgments}


\bibliographystyle{apsrev}

\begin{thebibliography}{10}
\bibitem{lambda_note} The penetration depth is defined here simply as the characteristic decay length of the magnetic field away from the vortex core. It cannot be defined via a London equation in the boundary theory because our boundary condition $\partial_z A_\phi = 0$ leads to zero current in the boundary theory.
\end{thebibliography}

\end {document}